\documentclass[conference]{IEEEtran}
\IEEEoverridecommandlockouts
\usepackage{cite}
\usepackage{amsmath,amssymb,amsfonts}
\usepackage{algorithmic}
\usepackage{graphicx}
\usepackage{textcomp}
\usepackage{xcolor}
\def\BibTeX{{\rm B\kern-.05em{\sc i\kern-.025em b}\kern-.08em
    T\kern-.1667em\lower.7ex\hbox{E}\kern-.125emX}}

\usepackage{subfigure}        
\begin{document}

\title{Comments on ``Federated Learning with Differential Privacy: Algorithms and Performance Analysis"}

\author{\IEEEauthorblockN{1\textsuperscript{st} Mahtab Talaei}
\IEEEauthorblockA{\textit{Department of Electrical and Computer Engineering} \\
\textit{Isfahan University of Technology}\\
Isfahan, Iran \\
mtalaei@bu.edu\textsuperscript{1}}
\and
\IEEEauthorblockN{2\textsuperscript{nd} Iman Izadi}
\IEEEauthorblockA{\textit{Department of Electrical and Computer Engineering} \\
\textit{Isfahan University of Technology}\\
Isfahan, Iran \\
iman.izadi@iut.ac.ir}
}

\maketitle

\begin{abstract}
In a previous paper \cite{bib1}, the convergence performance of the proposed differential privacy algorithm in federated learning (FL), noising before model aggregation FL (NbAFL), was studied. However, the presented convergence upper bound of the NbAFL (Theorem 2) is incorrect.
A comment is made here to present the correct form of the convergence upper bound of NbAFL.
\end{abstract}

\begin{IEEEkeywords}
Federated Learning, Differential Privacy, Convergence Performance
\end{IEEEkeywords}

\footnote[1]{Mahtab Talaei was affiliated with the Department of Electrical and Computer Engineering at Isfahan University of Technology during the research for this paper. At the time of submission, she is affiliated with the Division of Systems Engineering at Boston University, Boston, USA.}

\section{Introduction}
In \cite{bib1}, according to Theorem 2, the convergence upper bound of the NbAFL algorithm with protection level $\epsilon$ after $T$ aggregation rounds is given by
\begin{multline}
\mathbb{E}\lbrace{F(\tilde{w}^{T}) - F(w^{*})}\rbrace \leqslant \\ P^{T} \Theta + \left( \frac{k_{1} T}{\epsilon} + \frac{k_{0} T^{2}}{\epsilon ^{2}} \right) \left( 1 - P ^{T} \right),
\end{multline}
where by defining $\lambda_{0}=\frac{\rho}{2}, \lambda _{1} = 1 + \frac{\rho B}{\mu}$ (in the paper mistakenly stated as $\lambda _{1} = \frac{1}{\mu} + \frac{\rho B}{\mu}$), $\text{ and }  \lambda_{2}=-\frac{1}{\mu} + \frac{\rho B}{\mu ^{2}} + \frac{\rho B}{2 \mu ^{2}}$, we have $P= 1 + 2l \lambda _ {2}, k_{1}= \frac{\lambda_{1} \beta C c}{m (1-P)}\sqrt{\frac{2}{N\pi}},
\text{ and } k_0 = \frac{\lambda_{0} C^{2} c^{2}}{m^{2}(1-P)N}$.
\\

However, due to a misuse of the Polyak-Lojasiewicz inequality in the given proof (Appendix E), the right hand side of this inequality is not correct. The Polyak-Lojasiewicz inequality with positive parameter $l$ implies that
\begin{equation}
\mathbb{E}\lbrace F(\tilde{w}^{(t)}) - F(w^{*}) \rbrace \leqslant \frac{1}{2l} \Vert {\nabla F(\tilde{w}^{(t)})\Vert}^2. \label{eq:2}
\end{equation}
Moreover, based on Lemma 2 the upper bound of the expected increment in the loss function is calculated as
\begin{multline}
\mathbb{E}\lbrace F(\tilde{w}^{(t+1)}) -F(\tilde{w}^{(t)}) \rbrace \leqslant 
\lambda_{2} \mathbb{E}\lbrace \Vert \nabla F(\tilde{w}^{(t)})\Vert ^2 \rbrace \\
+ \lambda_{1} \mathbb{E}\lbrace \Vert n^{(t+1)} \Vert \Vert \nabla F(\tilde{w}^{(t)})\Vert \rbrace 
+ \lambda_{0} \mathbb{E}\lbrace \Vert n^{(t+1)} \Vert ^{2} \rbrace. \label{eq:3}
\end{multline}
Adding $\mathbb{E} \lbrace F(\tilde{w}^{(t)}) - F(w^{*}) \rbrace$ to both sides of \eqref{eq:3}, we have
\begin{multline}
\mathbb{E}\lbrace{F(\tilde{w}^{t+1}) - F(w^{*})}\rbrace \leqslant \\
\mathbb{E} \lbrace F(\tilde{w}^{(t)}) - F(w^{*}) \rbrace 
+ \lambda_{2} \mathbb{E}\lbrace \Vert \nabla F(\tilde{w}^{(t)})\Vert ^2 \rbrace \\
+ \lambda_{1} \mathbb{E}\lbrace \Vert n^{(t+1)} \Vert \Vert \nabla F(\tilde{w}^{(t)})\Vert \rbrace 
+ \lambda_{0} \mathbb{E}\lbrace \Vert n^{(t+1)} \Vert ^{2} \rbrace. \label{eq:4}
\end{multline}
Considering $\Vert \nabla F(\tilde{w}^{(t)})\Vert  \leqslant \beta$ , \cite{bib1} used inequality \eqref{eq:2} in an opposite direction and incorrectly obtained (50) in the paper as
\begin{multline}
\mathbb{E}\lbrace{F(\tilde{w}^{(t+1)}) - F(w^{*})}\rbrace \leqslant 
(1+ 2l \lambda_{2})\mathbb{E}\lbrace F(\tilde{w}^{(t)})- F(w^{*}) \rbrace \\
+ \lambda_{1} \beta \mathbb{E}\lbrace \Vert n^{(t+1)} \Vert \rbrace 
+ \lambda_{0} \mathbb{E}\lbrace \Vert n^{(t+1)} \Vert ^{2} \rbrace. \label{eq:5}
\end{multline}
In other words, while the right hand side phrases in \eqref{eq:4} can only be replaced with higher or equal ones, $\mathbb{E}\lbrace \Vert \nabla F(\tilde{w}^{(t)})\Vert ^2 \rbrace$ in \eqref{eq:4} is erroneously replaced by $2l \mathbb{E} \lbrace F(\tilde{w}^{(t)}) - F(w^{*}) \rbrace$ which is lower or equal than $\mathbb{E}\lbrace \Vert \nabla F(\tilde{w}^{(t)})\Vert ^2 \rbrace$.

To correct the right hand side of \eqref{eq:5}, from \eqref{eq:4} and without using Polyak-Lojasiewicz, we have
\begin{multline}
\mathbb{E}\lbrace{F(\tilde{w}^{(t+1)}) - F(w^{*})}\rbrace \leqslant \mathbb{E}\lbrace F(\tilde{w}^{(t)})- F(w^{*}) \rbrace  \\ + \lambda_{2} \beta ^{2} 
+ \lambda_{1} \beta \mathbb{E}\lbrace \Vert n^{(t+1)} \Vert \rbrace 
+ \lambda_{0} \mathbb{E}\lbrace \Vert n^{(t+1)} \Vert ^{2} \rbrace, \label{eq:6}
\end{multline}
where $F(w^{*})$ is the loss function corresponding to the optimal parameter $w^*$. Defining $\mathbb{E}\lbrace \Vert n^{(t)} \Vert \rbrace = \mathbb{E}\lbrace \Vert n \Vert \rbrace  \text{ and }  \mathbb{E}\lbrace \Vert n^{(t)} \Vert ^{2} \rbrace = \mathbb{E}\lbrace \Vert n \Vert ^2 \rbrace$, and applying \eqref{eq:6} recursively for $0 \leqslant t \leqslant T$ yields
\begin{multline}
\mathbb{E}\lbrace{F(\tilde{w}^{(T)}) - F(w^{*})}\rbrace \leqslant \mathbb{E}\lbrace F(\tilde{w}^{(0)})- F(w^{*}) \rbrace + \\  T \lambda_{2} \beta ^{2} 
+ T \lambda_{1} \beta \mathbb{E}\lbrace \Vert n \Vert \rbrace 
+ T \lambda_{0} \mathbb{E}\lbrace \Vert n \Vert ^{2} \rbrace, \label{eq:7}
\end{multline}
Substituting
$\mathbb{E}\lbrace \Vert n \Vert \rbrace = \frac{\Delta s _{D} Tc}{\epsilon}\sqrt{\frac{2N}{\pi}} \text{ and } 
\mathbb{E}\lbrace \Vert n \Vert ^2 \rbrace = \frac{\Delta s _{D} ^2 T ^2 c ^2 N}{\epsilon ^2}$ (from (52) in the paper) into \eqref{eq:7},
by setting $\Delta s_{D} = \frac{2C}{mN} \text{ and } F(\tilde{w}^{(T)}) - F(w^{*})=\Theta$ we have
\begin{equation}
\begin{split}
\mathbb{E}\lbrace & {F(\tilde{w}^{(T)}) - F(w^{*})} \rbrace  \leqslant \Theta + \lambda_{2} T \beta ^{2} \\
& + \frac{2 \lambda_{1} T ^2 \beta C c}{m \epsilon} \sqrt{\frac{2}{ N \pi}} 
+ \frac{4 \lambda_{0} T^3 C^2 c^2 }{m^2 \epsilon ^2 N} \\
& = \Theta + k_2 T + \frac{k_1 T^{2}}{\epsilon} + \frac{k_0 T^{3}}{\epsilon ^2},  
\end{split}\label{eq:8}
\end{equation}
where $k_{2} = \lambda_{2} \beta^{2}, k_{1}= \frac{2 \lambda_{1} \beta C c}{m} \sqrt{\frac{2}{N\pi}}, \text{ and } k_{0} = \frac{4 \lambda_{0} C^{2} c^{2}}{m^{2}N}$.
This completes the proof and gives a new correct upper bound.

From another perspective, the same result can directly be obtained from \eqref{eq:3}. Applying \eqref{eq:3} recursively for $0 \leqslant t \leqslant T$ yields
\begin{multline}
\mathbb{E}\lbrace{F(\tilde{w}^{(T)}) - F(\tilde{w}^{(0)}}\rbrace \leqslant T \lambda_{2} \beta ^{2} \\
+ T \frac{2 \lambda_{1}\beta T C c}{m \epsilon} \sqrt{\frac{2}{ N \pi}} 
+ T \frac{4 \lambda_{0} T^2 C^2 c^2 }{m^2 \epsilon ^2 N}.  \label{eq:9}
\end{multline}
Adding  $\mathbb{E} \lbrace F(\tilde{w}^{(0)}) - F(w^{*}) \rbrace$ to both sides of \eqref{eq:9}, we have
\begin{multline}
\mathbb{E}\lbrace{F(\tilde{w}^{(T)}) - F(w^{*})}\rbrace \leqslant \Theta + \lambda_{2} T \beta ^{2} \\
+ \frac{2 \lambda_{1} T ^2 \beta C c}{m \epsilon} \sqrt{\frac{2}{ N \pi}} 
+ \frac{4 \lambda_{0} T^3 C^2 c^2 }{m^2 \epsilon ^2 N}, \label{eq:10}
\end{multline}
which completes the proof.

\bibliographystyle{IEEEtran}
\bibliography{IEEEabrv,mybibfile}

\end{document}